\documentclass[a4paper]{jpconf}
\usepackage{graphicx}
\usepackage{amsmath,amssymb}
\usepackage{soul,color}
\begin{document}
\title{GJMS-like operators on symmetric 2-tensors and their gravitational duals}

\author{R Aros$^1$, F Bugini$^2$ and D E Diaz$^3$}

\address{$^1$ Departamento de Ciencias Fisicas, Universidad Andres Bello,
Sazie 2212, Piso 7, Santiago, Chile}
\address{$^2$ Departamento de Matem\'atica y F\'isica Aplicadas,
Universidad Cat\'olica de la Sant\'isima Concepci\'on,
Alonso de Ribera 2850, Concepci\'on, Chile}
\address{$^3$ Departamento de Ciencias Fisicas, Universidad Andres Bello,
Autopista Concepcion-Talcahuano 7100, Talcahuano, Chile}
\ead{raros@unab.cl, fbugini@ucsc.cl,danilodiaz@unab.cl}
\begin{abstract}
We study a family of higher-derivative conformal operators $P_{2k}^{(2)}$ acting on transverse-traceless symmetric 2-tensors on generic Einstein spaces. They are a natural generalization of the well-known construction for scalars.

We first provide the alternative description in terms of a bulk Poincar\'e-Einstein metric by making use of the AdS/CFT dictionary and argue that their holographic dual generically consists of bulk massive gravitons. 
At one-loop quantum level we put forward a holographic formula for the functional determinant of the higher-derivative conformal operators $P_{2k}^{(2)}$ in terms of the functional determinant for massive gravitons with standard and alternate boundary conditions. The analogous construction for vectors $P_{2k}^{(1)}$ is worked out as well and we also rewrite the holographic formula for unconstrained vector and traceless symmetric 2-tensor by decoupling the longitudinal part.      

Finally, we show that the holographic formula provides the necessary building blocks to address the massless and partially massless bulk gravitons. This is confirmed in four and six dimensions,  verifying full agreement with results available in the literature.

\end{abstract}

\vspace{20mm}

\section{Introduction}

In this note, we elaborate on a family of higher-derivative conformal operators $P_{2k}^{(2)}$ acting on transverse-traceless 2-tensors. In the Fefferman-Graham approach to conformal geometry, they are derived from powers of the Lichnerowicz Laplacian $\tilde{\Delta}_L^k$ of the ambient Lorentzian metric. Although obstructed in general, on even-dimensional Einstein manifolds they happen to exist and, furthermore, they factorize into products of the boundary Lichnerowicz Laplacian as first derived in \cite{Matsumoto_2013}
\begin{equation}
\displaystyle P_{2k}^{(2)}=\prod_{j=1}^{k}\left\{\Delta_L^{(2)}-4(n-1)\lambda +2\left(\frac{n}{2}-j\right)\left(j+\frac{n}{2}-1\right)\lambda\right\}~,
\end{equation}
where $n$ is the dimension of the boundary Einstein manifold and $\lambda$ stands for $n$ times the trace of the Schouten tensor\footnote{The Schouten tensor in terms of the Ricci tensor and Ricci scalar is given by $P_{ij}=(R_{ij}-R/(2(n-1))g_{ij})/(n-2)$, and for an Einstein manifold is proportional to the metric $P_{ij}=\lambda g_{ij}$ with $n\lambda
=P^i_i=R/(2(n-1))$.}, proportional to the necessarily constant Ricci scalar $R$.

Our aim is to put these GJMS-like operators\footnote{The superscripts (0), (1) and (2) on the GJMS-like operators $P_{2k}$ and on the Lichnerowicz Laplacians $\Delta_L$ refer to their action on scalar, vector and symmetric two-tensor, respectively.} on equal footing with the original scalar GJMS operators $P_{2k}^{(0)}$. This will be done by finding out the appropriate extension of the following two central features regarding the holographic counterpart in a Poincar\'e-Einstein bulk metric:
\begin{enumerate}
    \item[(i)] at tree level the scalar GJMS operators are obtained from a bulk massive scalar with mass $m^2=k^2-\frac{n^2}{4}$
\begin{equation}
\left\{\hat{\Delta}_L^{(0)}+m^2\right\}\varphi=0
\end{equation}\\

\item[(ii)] at one-loop level there is a holographic formula relating the functional determinants

\begin{equation}
\displaystyle
\frac{{\det}_{-}\left\{\hat{\Delta}_L^{(0)}-\frac{n^2}{4}+k^2\right\}}{{\det}_{+}\left\{\hat{\Delta}_L^{(0)}-\frac{n^2}{4}+k^2\right\}}  = {\det}\,P_{2k}^{(0)}~,
\end{equation}
where the bulk $\pm$-determinants are computed with standard and alternate boundary conditions~(see e.g. \cite{Giombi:2013yva}).
\end{enumerate}

The organization of this paper is as follows. In section 2 we review the ambient construction and derive the explicit factorized form of the family of higher-derivative conformal operators $P_{2k}^{(1)}$ acting on transverse vectors.
We then provide in section 3 the alternative description in terms of a bulk Poincar\'e-Einstein metric by making use of the AdS/CFT dictionary and examine the particular mass values for which the bulk fields acquire an additional gauge invariance. 
In section 4  we move on to the one-loop quantum level and put forward a holographic formula for the functional determinant of the higher-derivative conformal operators $P_{2k}^{(2)}$ in terms of the functional determinant for massive gravitons with standard and alternate boundary conditions. In the process, the analogous construction for vectors $P_{2k}^{(1)}$ is worked out as well and we end up with an interesting recursive structure. 
In section 5 we rewrite the holographic formula for unconstrained vector and traceless symmetric 2-tensor by decoupling the longitudinal part.      
Finally, in section 6 we provide evidence for the correctness of the holographic formula by computing partition functions and Weyl anomaly coefficients in four and six dimensions, verifying full agreement with results available in the literature. The boundary values of the anomaly coefficients are collected in Appendix A.

\section{Ambient construction }

For completeness, let us start by recreating Matsumoto's derivation \cite{Matsumoto_2013} in terms of the ambient metric, adapted to the vector field. The ambient metric $\tilde{g}$, having an Einstein representative $g$ in the conformal class of boundary metrics, is a Ricci flat Lorentzian metric given by

\begin{equation}
  \tilde{g}=2\rho dt^2 + 2tdtd\rho+t^2\left(1+\lambda\rho\right)^2g~.
\end{equation}
The extra directions $t$ and $\rho$ are usually termed $0$ and $\infty$ respectively. Again $\lambda$ stands for $n$ times the trace of the Schouten tensor. From now on tildes, $\sim$, will denote the structures, and their components, in the ambient space. By the same token, hats, $\wedge$, will denote the structures and components in the bulk Poincar\'e-Einstein metric. 

The ambient inverse metric is then
\vspace{2mm}
\begin{equation}
    \tilde{g}^{\hspace{1ex}IJ}=
    \begin{pmatrix}
    &0 & & 0 & & t^{-1}&\\
    & & &  & & &\\
    &0 & & t^{-2}\left(1+\lambda \rho\right)^{-2}g^{ij} && 0& \\
    & & &  & & &\\
    &t^{-1} & & 0& & -2t^{-2}\rho&
    \end{pmatrix}
\end{equation}
\vspace{2mm}
with nonvanishing Christoffel symbols
\vspace{2mm}
\begin{equation}
    \tilde{\Gamma}^{\hspace{1ex}0}_{\hspace{2ex}IJ}=
    \begin{pmatrix}
    &0 & & 0 & & 0&\\
    & & &  & & &\\
    &0 & & -\lambda t \left(1+\lambda \rho\right)g_{ij} && 0& \\
    & & &  & & &\\
    &0 & & 0& & 0&
    \end{pmatrix}
\end{equation}
\vspace{2mm}

\begin{equation}
    \tilde{\Gamma}^{\hspace{1ex}k}_{\hspace{2ex}IJ}=
    \begin{pmatrix}
    &0 & & t^{-1}\delta_j^k & & 0&\\
    & & &  & & &\\
    &t^{-1}\delta_i^k & & \Gamma^{\hspace{1ex}k}_{\hspace{2ex}ij} && \lambda \left(1+\lambda \rho\right)^{-1}\delta_i^k& \\
    & & &  & & &\\
    &0 & & \lambda \left(1+\lambda \rho\right)^{-1}\delta_j^k& & 0&
    \end{pmatrix}
\end{equation}
\vspace{2mm}

\begin{equation}
    \tilde{\Gamma}^{\hspace{1ex}\infty}_{\hspace{2ex}IJ}=
    \begin{pmatrix}
    &0 & & 0& & t^{-1}&\\
    & & &  & & &\\
    &0 & & -\left(1-\lambda^2 \rho^2\right)g_{ij} && 0& \\
    & & &  & & &\\
    &t^{-1} & & 0& & 0&
    \end{pmatrix}
\end{equation}
\vspace{2mm}
In order to build up the ambient Laplacian on an ambient vector section of weight $w$
\begin{equation}
\tilde{\sigma}_i=t^w(1+\lambda\rho)^w\sigma_i \quad,\quad \tilde{\sigma}_0=\tilde{\sigma}_{\infty}=0
\end{equation}
we need the first derivatives
\begin{eqnarray}
\widetilde{\nabla}_{\infty} \tilde{\sigma}_{i}  &=& \partial_{\rho} \tilde{\sigma}_{i} - \tilde{\Gamma}^{ k}_{\hspace{1ex}\infty i}\,\sigma_k=t^w(1+\lambda\rho)^{w-1}(w-1)\lambda\sigma_i \nonumber\\
\nonumber\\
\widetilde{\nabla}_{0} \tilde{\sigma}_{i}  &=& {\partial}_{t} \tilde{\sigma}_{i} - \tilde{\Gamma}^{k}_{\hspace{1ex}0 i}\,\sigma_k=t^{w-1}(1+\lambda\rho)^w(w-1)\sigma_i \nonumber\\
\nonumber\\
\widetilde{\nabla}_{k} \tilde{\sigma}_{i}  &=& {\partial}_{k} \tilde{\sigma}_{i} - \tilde{\Gamma}^{l}_{\hspace{1ex}k i}\,\sigma_l=t^w(1+\lambda\rho)^{w}\nabla_k\sigma_i\\
\nonumber\\
\widetilde{\nabla}_{k} \tilde{\sigma}_{\infty}  &=& - \tilde{\Gamma}^{l}_{\hspace{1ex}k \infty}\,\sigma_l=-t^w(1+\lambda\rho)^{w-1}\lambda\sigma_k \nonumber\\
\nonumber\\
\widetilde{\nabla}_{k} \tilde{\sigma}_{0}  &=& - \tilde{\Gamma}^{l}_{\hspace{1ex}k 0}\,\sigma_l=-t^{w-1}(1+\lambda\rho)^{w}\sigma_k \nonumber
\end{eqnarray}
and then the non-vanishing second derivatives
\begin{eqnarray}
\widetilde{\nabla}_{0}\widetilde{\nabla}_{\infty} \tilde{\sigma}_{i}  &=& \partial_{t} \widetilde{\nabla}_{\infty}\tilde{\sigma}_{i} - \tilde{\Gamma}^{ \infty}_{\hspace{1ex}0 \infty}\widetilde{\nabla}_{\infty}\tilde{\sigma}_i - \tilde{\Gamma}^{ k}_{\hspace{1ex}0 i}\widetilde{\nabla}_{\infty}\tilde{\sigma}_k \nonumber\\
&=& t^{w-1}(1+\lambda\rho)^{w-1}(w-1)(w-2)\lambda\sigma_i \nonumber\\
& & \nonumber\\
\widetilde{\nabla}_{\infty}\widetilde{\nabla}_{0} \tilde{\sigma}_{i}  &=& \widetilde{\nabla}_{0}\widetilde{\nabla}_{\infty} \tilde{\sigma}_{i} - \tilde{R}_{\infty 0 \hspace{1ex} i}^{\hspace{3ex} k}
\,\tilde{\sigma}_k \nonumber \\
&=&\widetilde{\nabla}_{0}\widetilde{\nabla}_{\infty} \tilde{\sigma}_{i} \\
& & \nonumber \\
\widetilde{\nabla}_{\infty}\widetilde{\nabla}_{\infty} \tilde{\sigma}_{i}  &=& {\partial}_{\rho} \widetilde{\nabla}_{\infty} \tilde{\sigma}_{i} - \tilde{\Gamma}^{k}_{\hspace{1ex}\infty  i}\widetilde{\nabla}_{\infty} \tilde{\sigma}_{k} \nonumber \\
&=& t^w(1+\lambda\rho)^{w-2}(w-1)(w-2)\lambda^2\sigma_i \nonumber\\
& & \nonumber \\
g^{kl}\widetilde{\nabla}_{k}\widetilde{\nabla}_{l} \tilde{\sigma}_{i}  &=& g^{kl}\left\{\partial_k \widetilde{\nabla}_{l} \tilde{\sigma}_{i} - \tilde{\Gamma}^{m}_{\hspace{1ex}k i }\widetilde{\nabla}_{l}\tilde{\sigma}_m - \tilde{\Gamma}^{\infty}_{\hspace{1ex}k l }\widetilde{\nabla}_{\infty}\tilde{\sigma}_i - \tilde{\Gamma}^{\infty}_{\hspace{1ex} k i }\widetilde{\nabla}_{l}\tilde{\sigma}_{\infty}- \tilde{\Gamma}^{0}_{\hspace{1ex} k l }\widetilde{\nabla}_{0}\tilde{\sigma}_i - \tilde{\Gamma}^{0}_{\hspace{1ex}k i }\widetilde{\nabla}_{l}\tilde{\sigma}_0\right\} \nonumber\\ &=&-t^w(1+\lambda\rho)^{w}\left\{\Delta+2\lambda-2(w-1)n\lambda\right\}\sigma_i. \nonumber
\end{eqnarray}
The ambient Laplacian on the vector section is then given by
\begin{eqnarray}
\widetilde{\Delta}\,\tilde{\sigma}_{i} &=& -2t^{-1}\widetilde{\nabla}_{0}\widetilde{\nabla}_{\infty} \tilde{\sigma}_{i}+ 2t^{-2}\rho\widetilde{\nabla}_{\infty}\widetilde{\nabla}_{\infty} \tilde{\sigma}_{i} -t^{-2}(1+\rho)^{-2}g^{kl}\widetilde{\nabla}_{k}\widetilde{\nabla}_{l} \tilde{\sigma}_{i}\\
&=&t^{w-2}(1+\lambda\rho)^{w-2}\left\{\Delta+2\lambda-2(w-1)(n+w-2)\lambda\right\}\sigma_i~. \nonumber
\end{eqnarray}
Finally, acting $k$-times on vectors of weight $w=-\frac{n}{2}+k+1$ yields the factorized form of the GJMS-like operators in terms of the Lichnerowicz Laplacian $\Delta_L^{(1)}=\Delta+2(n-1)\lambda$ as follows
\begin{equation}
\displaystyle P_{2k}^{(1)}=\prod_{j=1}^{k}\left\{\Delta_L^{(1)}-2(n-2)\lambda +2\left(\frac{n}{2}-j\right)\left(j+\frac{n}{2}-1\right)\lambda\right\}~.
\end{equation}

This formula comprises several instances of conformal operators already reported in the literature. The $k=1$ representative corresponds to the conformal two-derivative vector of Erdmenger and Osborn \cite{Erdmenger:1997wy}. In general dimensions, it describes a conformal but non-gauge vector, except in 4D where it becomes the Maxwell field with the additional gauge invariance.  The $k=2$ case corresponds to a 4-derivative conformal gauge vector in 6D. Massive representations in 6D bear the same form as in the 6-sphere (c.f. eqn.A.17 in \cite{Beccaria:2015uta} and Eqn.C.4 in \cite{Beccaria:2017dmw}).

\section{Tree Level}

The alternative description in terms of the bulk Poincar\'e-Einstein metric by making use of the AdS/CFT dictionary can be inferred from the one in Euclidean $AdS_{n+1}$ or hyperbolic space~\cite{Giombi:2013yva}. Symmetric transverse-traceless 2-tensors satisfy the Fierz-Pauli equation with mass $m^2=k^2-\frac{n^2}{4}$
\begin{equation}
\left\{\hat{\Delta}_L^{(2)}+2n+m^2\right\}\varphi_{_{\bot\top}}=0,
\end{equation}
whereas transverse vectors satisfy the Proca equation with mass $m^2=k^2-\frac{(n-2)^2}{4}$
\begin{equation}
\left\{\hat{\Delta}_L^{(1)}+m^2\right\}\varphi_{_{\bot}}=0~.
\end{equation}

An unconstrained symmetric 2-tensor in $n+1$ dimensions has $(n+1)(n+2)/2$ independent components; whereas for a transverse-traceless 2-tensor, after subtracting one component of the trace scalar and $n+1$ of the longitudinal part, one ends up with $(n-1)(n+2)/2$ components. The boundary dual of the bulk transverse-traceless 2-tensor corresponds to a traceless 2-tensor with $n(n+1)/2-1=(n-1)(n+2)/2$ components.

For special values of the mass, however, some degrees of freedom become redundant due to gauge invariance. Below we discuss the two possible gauge invariances for the rank-2 symmetric tensor.

\subsection{Vector gauge invariance: bulk Einstein graviton}

The massless bulk field has $m^2=0$ and, therefore, $k=n/2$ and kinetic term $\left\{\hat{\Delta}_L^{(2)}+2n\right\}$. The corresponding vector ghost has the very same form in terms of the Lichnerowicz Laplacian $\left\{\hat{\Delta}_L^{(1)}+2n\right\}$ and its mass being then $m^2=k'^2-(n-2)^2/4=2n$. It corresponds therefore to $k'=n/2+1$.

\subsection{Scalar gauge invariance: bulk partially massless graviton}

The partially massless bulk field has $m^2=1-n$ (see e.g.  eqn.(3.20) in \cite{Tseytlin:2013jya}) and, therefore, $k=n/2-1$. Thus, the kinetic term is given by $\left\{\hat{\Delta}_L^{(2)}+n+1\right\}$. The corresponding scalar ghost has again the very same form in terms of the Lichnerowicz Laplacian $\left\{\hat{\Delta}_L^{(0)}+n+1\right\}$ with $m^2=k'^2-n^2/4=n+1$. It corresponds therefore to $k'=n/2+1$.

\section{One loop: functional determinants}
We propose the following natural extension of the holographic formula to transverse vectors
\vspace{2mm}
\begin{equation}
\displaystyle
\frac{{\det}_{-,_{\bot}}\left\{\hat{\Delta}_L^{(1)} -\frac{(n-2)^2}{4} + k^2\right\}}{{\det}_{+,_{\bot}}\left\{\hat{\Delta}_L^{(1)} - \frac{(n-2)^2}{4} + k^2\right\}}  ={\det}_{_{\bot}}\,P_{2k}^{(1)}\cdot{\det}\,P_{2k}^{(0)}={\det}\,P_{2k}^{(1)}
\end{equation}
and for symmetric transverse-traceless 2-tensors
\begin{equation}
\displaystyle \frac{{\det}_{-,_{\bot\top}}\left\{\hat{\Delta}_L^{(2)}+2n-\frac{n^2}{4}+k^2\right\}}{{\det}_{+,_{\bot\top}}\left\{\hat{\Delta}_L^{(2)}+2n-\frac{n^2}{4}+k^2\right\}}  ={\det}_{_{\bot\top}}\,P_{2k}^{(2)}\cdot{\det}_{_{\bot}}\,P_{2k}^{(1)}\cdot{\det}\,P_{2k}^{(0)}={\det}_{_{\top}}\,P_{2k}^{(2)}~.
\end{equation}
The bulk determinants are to be computed in the space-filling Poincar\'e-Einstein metric with an Einstein metric on the conformal infinity, whereas the boundary determinants are computed on the boundary Einstein manifold. In the  last equality in each formula, we have absorbed the longitudinal part of the vector and traceless symmetric 2-tensor, respectively. The explicit expressions for the resulting non-minimal operators will be given in the next section.

It is worth noticing that the conformal nature of the boundary functional determinants becomes explicit by the presence of the GJMS-like operators, a feature that is not apparent when written in terms of the various Lichnerowicz Laplacians that enter the factorizations.

These formulas were meant to hold just for massive bulk fields, but they can be conjectured to serve as building blocks in the case of massless and partially-massless bulk fields. This will be illustrated by the explicit computation of the central charges in 4D and 6D, see below in section 6.

\subsection{5D bulk Einstein graviton / 4D boundary Weyl graviton}

The massless bulk Einstein graviton corresponds to $k=n/2=2$ in 4D and must be accompanied by a vector ghost contribution
\begin{equation}
\displaystyle \frac{{\det}_{_{\bot\top}}\left\{\hat{\Delta}_L^{(2)}+8\right\}}{{\det}_{_{\bot}}\left\{\hat{\Delta}_L^{(1)}+8\right\}}~.
\end{equation}
As explained before, the ghost determinant corresponds to a massive vector with $k'=n/2+1=3$, so the following quotient is obtained for the boundary determinants upon application of the proposed holographic formulas
\begin{equation}
\displaystyle
\frac{{\det}_{_{\bot\top}}\,P_{4}^{(2)}\cdot{\det}_{_{\bot}}\,P_{4}^{(1)}\cdot{\det}\,P_{4}^{(0)}}{{\det}_{_{\bot}}\,P_{6}^{(1)}\cdot{\det}\,P_{6}^{(0)}}.
\end{equation}
Inserting now the explicit factorized form for the GJMS-like conformal operators
\begin{eqnarray}
P_4^{(2)}&=&
\left\{\Delta_L^{(2)}-6\right\}\cdot\left\{\Delta_L^{(2)}-4\right\}
\\\nonumber\\
P_6^{(1)}&=&P_4^{(1)}\cdot\left\{\Delta_L^{(1)}-6\right\}
\\\nonumber\\
P_6^{(0)}&=&P_4^{(0)}\cdot\left\{\Delta_L^{(0)}-4\right\}
\end{eqnarray}
we correctly reproduce the one-loop partition function for the 4D Weyl graviton \cite{Tseytlin:2013jya}
\begin{equation}
\displaystyle
Z_{_{Weyl}}^{^{1-loop}}=\left\{\frac{{\det}_{_{\bot\top}}\,\{\Delta_L^{(2)}-6\}}{{\det}_{_{\bot}}\,\{\Delta_L^{(1)}-6\}}\cdot \frac{{\det}_{_{\bot\top}}\,\{\Delta_L^{(2)}-4\}}{{\det}\,\{\Delta_L^{(0)}-4\}}\right\}^{-1/2}~.
\end{equation}
The first quotient is due to the boundary Einstein graviton which is non-conformal but it has vector gauge invariance; the second quotient, in turn, is due to the boundary partially-massless graviton which in 4D is conformal and has scalar gauge invariance.

\subsection{5D bulk partially massless graviton/4D boundary conformal symmetric tensor}

For the partially massless bulk graviton, we had $k=n/2-1=1$ in 4D and must be accompanied by a scalar ghost contribution
\begin{equation}
\displaystyle \frac{{\det}_{_{\bot\top}}\left\{\hat{\Delta}_L^{(2)}+5\right\}}{{\det}\left\{\hat{\Delta}_L^{(0)}+5\right\}}~.
\end{equation}
As explained before, the ghost determinant corresponds to a massive scalar with $k'=n/2+1=3$, so the following quotient is obtained for the boundary determinants upon application of the holographic formulas
\begin{equation}
\displaystyle
\frac{{\det}_{_{\bot\top}}\,P_{2}^{(2)}\cdot{\det}_{_{\bot}}\,P_{2}^{(1)}\cdot{\det}\,P_{2}^{(0)}}{{\det}\,P_{6}^{(0)}}.
\end{equation}
Inserting now the factorized form for the GJMS-like conformal operators in terms of Lichnerowicz Laplacians
\begin{eqnarray}
P_2^{(2)}&=&\left\{\Delta_L^{(2)}-4\right\}
\\\nonumber\\
P_2^{(1)}&=&\left\{\Delta_L^{(1)}\right\}
\\\nonumber\\
P_6^{(0)}&=&P_2^{(0)}\cdot\left\{\Delta_L^{(0)}-4\right\}\cdot\left\{\Delta_L^{(0)}\right\}
\end{eqnarray}
we obtain the partition function for the boundary dual of the bulk partially massless graviton. This corresponds to the one-loop partition function of the \textit{ conformal symmetric tensor} discussed in \cite{Beccaria:2015vaa}  \footnote{Our result is valid for a generic Einstein boundary and contains the two cases considered in \cite{Beccaria:2015vaa}, eqn.3.16 on Ricci flat and eqn.3.19 on the 4-sphere therein. See also \cite{Kuzenko:2019eni}, eqn.3.13.}
\begin{equation}
\displaystyle
Z_{_{CST}}^{^{1-loop}}=\left\{\frac{{\det}_{_{\bot\top}}\,\{\Delta_L^{(2)}-4\}}{{\det}\,\{\Delta_L^{(0)}-4\}}\cdot \frac{{\det}_{_{\bot}}\,\{\Delta_L^{(1)}\}}{{\det}\,\{\Delta_L^{(0)}\}}\right\}^{-1/2}~.
\end{equation}
The first quotient, already identified, is due to the boundary partially-massless graviton which in 4D is conformal and has scalar gauge invariance; while the second quotient in due to the boundary Maxwell field which is both conformal and scalar gauge invariant.

\subsection{7D bulk Einstein graviton / 6D boundary Weyl graviton}

The massless bulk Einstein graviton now corresponds to $k=n/2=3$ in 6D accompanied by the vector ghost contribution
\begin{equation}
\displaystyle \frac{{\det}_{_{\bot\top}}\left\{\hat{\Delta}_L^{(2)}+12\right\}}{{\det}_{_{\bot}}\left\{\hat{\Delta}_L^{(1)}+12\right\}}.
\end{equation}
The ghost determinant corresponds to a massive vector with $k'=n/2+1=4$, so that the quotient for the boundary determinants becomes
\begin{equation}
\displaystyle
\frac{{\det}_{_{\bot\top}}\,P_{6}^{(2)}\cdot{\det}_{_{\bot}}\,P_{6}^{(1)}\cdot{\det}\,P_{6}^{(0)}}{{\det}_{_{\bot}}\,P_{8}^{(1)}\cdot{\det}\,P_{8}^{(0)}}~.
\end{equation}

By using the factorized form for the GJMS-like conformal operators in terms of Lichnerowicz Laplacians
\begin{eqnarray}
P_6^{(2)}&=&\left\{\Delta_L^{(2)}-10\right\}\cdot\left\{\Delta_L^{(2)}-6\right\}\cdot\left\{\Delta_L^{(2)}-4\right\}
\\\nonumber\\
P_8^{(1)}&=&P_6^{(1)}\cdot\left\{\Delta_L^{(1)}-10\right\}
\\\nonumber\\
P_8^{(0)}&=&P_6^{(0)}\cdot\left\{\Delta_L^{(0)}-6\right\}
\end{eqnarray}
we recover, as expected, the one-loop partition function for the 6D Weyl graviton discussed in detail in \cite{Aros_2019JHEP}
\begin{equation}
\displaystyle
Z_{_{Weyl}}^{^{1-loop}}=\left\{\frac{{\det}_{_{\bot\top}}\,\{\Delta_L^{(2)}-10\}}{{\det}_{_{\bot}}\,\{\Delta_L^{(1)}-10\}}\cdot \frac{{\det}_{_{\bot\top}}\,\{\Delta_L^{(2)}-6\}}{{\det}\,\{\Delta_L^{(0)}-6\}}\cdot {\det}_{_{\bot\top}}\,\{\Delta_L^{(2)}-4\}\right\}^{-1/2}~.
\end{equation}
The first quotient is due to the boundary Einstein graviton which is gauge invariant and non-conformal; the second quotient is due to the boundary partially massless graviton which is also gauge-invariant, with a scalar parameter, and non-conformal; and finally, the third quotient corresponds to a massive boundary graviton which is non-gauge but conformal in 6D, it actually corresponds to the rank-2 Erdmenger-Osborn transverse-traceless field  \cite{Erdmenger:1997wy}.

\subsection{7D bulk partially massless graviton / 6D boundary conformal symmetric tensor}

The partially massless bulk graviton now has $k=n/2-1=2$ in 6D accompanied by a scalar  ghost contribution
\begin{equation}
\displaystyle \frac{{\det}_{_{\bot\top}}\left\{\hat{\Delta}_L^{(2)}+7\right\}}{{\det}_{_{\bot}}\left\{\hat{\Delta}_L^{(1)}+7\right\}}.
\end{equation}
The ghost determinant corresponds now to a massive scalar with $k'=n/2+1=4$, so that the quotient for the boundary determinants becomes
\begin{equation}
\displaystyle
\frac{{\det}_{_{\bot\top}}\,P_{4}^{(2)}\cdot{\det}_{_{\bot}}\,P_{4}^{(1)}\cdot{\det}\,P_{4}^{(0)}}{\det P_{8}^{(0)}}.
\end{equation}
The factorized form for the GJMS-like conformal operators in terms of Lichnerowicz Laplacians
\begin{eqnarray}
P_4^{(2)}&=&\left\{\Delta_L^{(2)}-6\right\}\cdot\left\{\Delta_L^{(2)}-4\right\}
\\\nonumber\\
P_4^{(1)}&=&\left\{\Delta_L^{(1)}+2\right\}\cdot\left\{\Delta_L^{(1)}\right\}
\\\nonumber\\
P_8^{(0)}&=&P_4^{(0)}\cdot\left\{\Delta_L^{(0)}\right\}\cdot\left\{\Delta_L^{(0)}-6\right\}
\end{eqnarray}
allows to rewrite the partition function for the 6D analog of the aforementioned \textit{conformal symmetric tensor} as
\begin{equation}
\displaystyle
Z_{_{CST}}^{^{1-loop}}=\left\{ \frac{{\det}_{_{\bot\top}}\,\{\Delta_L^{(2)}-6\}}{{\det}\,\{\Delta_L^{(0)}-6\}}\cdot {\det}_{_{\bot\top}}\,\{\Delta_L^{(2)}-4\}\cdot\frac{{\det}_{_{\bot}}\,\{\Delta_L^{(1)}\}}{{\det}\,\{\Delta_L^{(0)}\}}\cdot {\det}_{_{\bot}}\,\{\Delta_L^{(1)}+2\}\right\}^{-1/2}~.
\end{equation}
The first quotient is due to the boundary partially massless graviton which is also gauge-invariant, with a scalar parameter, and non-conformal; the second factor corresponds, as before, to a massive boundary graviton which is non-gauge but conformal, corresponding to the rank-2 Erdmenger-Osborn transverse-traceless field \cite{Erdmenger:1997wy}; the second quotient corresponds to a 6D non-conformal vector with scalar gauge invariance (6D Maxwell field, c.f. eqn. A.13 in~\cite{Beccaria:2015uta}); and finally, the last determinant comes from a conformal and non-gauge 6D massive vector~\footnote{In fact, this determinant combined with the previous quotient produce the partition function for the 4-derivative conformal gauge vector which is the $s=1$ member of the Conformal Higher Spin (CHS) family in 6D (see eqn.A.11 in~\cite{Beccaria:2015uta}, and also eqn.5.25 in \cite{Mukherjee:2021alj}. Besides, the massive and the partially massless gravitons combine to form, on the six-sphere, a maximal depth $t=s=2$ CHS with residual scalar gauge invariance as discussed in eqn.3.26 of \cite{Grigoriev:2018mkp}.} (Erdmenger-Osborn conformal vector \cite{Erdmenger:1997wy}).

\section{One-loop: minimal vs. nonminimal operators}

Lets us now turn to the decoupling of the longitudinal part for the unconstrained vector and traceless symmetric 2-tensor on a generic Einstein background.

\subsection{Vector}
For the purpose of illustration, take the $k=1$ representative of the vector family that corresponds to the Erdmenger-Osborn\cite{Erdmenger:1997wy} operator, that we also denote by $P_2^{(1)}$ but acting on unconstrained vectors $A_{\mu}$
\begin{equation}
\displaystyle A^{\mu}\,P_2^{(1)}A_{\mu}\,=\,A^{\mu}\left\{\Delta_L^{(1)}+\frac{4}{n}\,\nabla\,\nabla\cdot\,+\,\frac{\lambda}{2}(n-2)(n-4)\right\}A_{\mu}~,
\end{equation}
and split into transverse and longitudinal components $A_{\mu}=A^{\bot}_{\mu}+\partial_{\mu}\varphi$. The intertwining properties of the Lichnerowicz Laplacian allow the diagonalization of the operator  
\begin{equation}
\displaystyle A^{\bot,\,\mu}\left\{\Delta_L^{(1)}+\,\frac{\lambda}{2}(n-2)(n-4)\right\}A^{\bot}_{\mu}\,+\,\frac{n-4}{n}\,\varphi\,\Delta_L^{(0)}\,\left\{\Delta_L^{(0)}\,+\,\frac{\lambda}{2}n(n-2)\right\}\varphi~,
\end{equation}
and one can readily identify the $k=1$ GJMS-like operators on the transverse vector and on the scalar~\footnote{The `accidental' value $n=4$ where the scalar goes away clearly produces the 4D Maxwell field which is both gauge and conformal invariant.}. The scalar is accompanied by an additional $\Delta_L^{(0)}$ factor that makes the operator quartic in derivatives, but it cancels out against the Jacobian of the splitting in the one-loop functional determinant so that one ends up with the following identity
\begin{equation}
\displaystyle
{\det}\,P_{2}^{(1)}\,=\,{\det}_{_{\bot}}\,P_{2}^{(1)}\cdot{\det}\,P_{2}^{(0)}~.
\end{equation}
The factorized structure of the GJMS-like operator on unconstrained vectors for generic $k$
on an Einstein manifold can be inferred from the conformally flat case\cite{Grigoriev:2018mkp}
\begin{equation}
\displaystyle P_{2k}^{(1)}=\prod_{j=1}^{k}\left\{\Delta_L^{(1)}+\frac{n-2}{(\frac{n}{2}-j)(\frac{n}{2}+j-1)}\,\nabla\,\nabla\cdot\,-\,2(n-2)\lambda +2\left(\frac{n}{2}-j\right)\left(j+\frac{n}{2}-1\right)\lambda\right\}~.
\end{equation}
For each factor, say j-block, one gets again a decoupling between transverse and longitudinal components with the j-blocks of the corresponding GJMS factors
\begin{eqnarray}
\displaystyle 
A^{\bot,\,\mu}\left\{\Delta_L^{(1)}-2(n-2)\lambda +2\left(\frac{n}{2}-j\right)\left(j+\frac{n}{2}-1\right)\lambda\right\}A^{\bot}_{\mu}+\\ \nonumber
\\\nonumber
+\,\left[1-\frac{n-2}{(\frac{n}{2}-j)(\frac{n}{2}+j-1)}\right]\,\varphi\,\Delta_L^{(0)}\,\left\{\Delta_L^{(0)}+2\left(\frac{n}{2}-j\right)\left(j+\frac{n}{2}-1\right)\lambda\right\}\varphi~.
\end{eqnarray}
The additional $\Delta_L^{(0)}$ factor again cancels out against the Jacobian  and one ends up with an identity valid now for generic order $k$
\begin{equation}
\displaystyle
{\det}\,P_{2k}^{(1)}\,=\,{\det}_{_{\bot}}\,P_{2k}^{(1)}\cdot{\det}\,P_{2k}^{(0)}~.
\end{equation}

\subsection{Symmetric 2-tensor}
Let us now consider the $k=1$ representative of the traceless symmetric 2-tensor family that corresponds to the Erdmenger-Osborn\cite{Erdmenger:1997wy} operator with a particular Lichnerowicz coupling~\footnote{The conformally invariant Erdmenger-Osborn operator has freedom in the coupling with the Weyl tensor. Our particular Lichnerowicz choice for the coupling will prove crucial in the decoupling of the transverse from the longitudinal components.} to the Weyl tensor, denoted by $P_2^{(2)}$ and acting on unconstrained traceless symmetric 2-tensors $\varphi_{_{\top},\mu\nu}$
\begin{equation}
\displaystyle \varphi_{_{\top}}^{\mu\nu}\,P_2^{(2)}\varphi_{_{\top},\mu\nu}\,=\,\varphi_{_{\top}}^{\mu\nu}\left\{\Delta_L^{(2)}+\frac{8}{n+2}\,\nabla\,\nabla\cdot\,+\,\frac{\lambda}{2}\left(n^2-10n+8\right)\right\}\varphi_{_{\top},\mu\nu}~.
\end{equation}
We split now into transverse and longitudinal components \begin{equation}
\displaystyle\varphi_{_{\top},\mu\nu}=\varphi_{_{\bot\top},\mu\nu}+\nabla^{\mu}V_{_{\bot}}^{\nu}+\nabla^{\nu}V_{_{\bot}}^{\mu}+(\nabla^{\mu}\nabla^{\nu}-\frac{g^{\mu\nu}}{n}\nabla^2)\,\sigma~.
\end{equation} 
The intertwining properties of the Lichnerowicz Laplacians allow again the diagonalization of the operator  
\begin{eqnarray}
\displaystyle &\varphi_{_{\bot\top}}^{\mu\nu}\left\{\Delta_L^{(2)}\,+\,\frac{\lambda}{2}\left(n^2-10n+8\right)\right\}\varphi_{_{\bot\top},\mu\nu}\,+\\\nonumber
&+\,2\,\frac{n-2}{n+2}\,V^{\bot,\,\mu}\left\{\Delta_L^{(1)}-4\lambda(n-1)\right\}\left\{\Delta_L^{(1)} +\frac{\lambda}{2}(n-2)(n-4)\right\}V^{\bot}_{\mu}+\\ \nonumber
\\
\nonumber
&+\frac{(n-1)(n-2)(n-4)}{n^2(n+2)}\,\sigma\,\Delta_L^{(0)}\,\left\{\Delta_L^{(0)}-2n\lambda\right\}\,\left\{\Delta_L^{(0)}+\frac{\lambda}{2}n(n-2)\right\}\sigma~.
\end{eqnarray}
The $k=1$ GJMS-like operators on the transverse-traceless symmetric 2-tensor, on the transverse vector, and on the scalar come out~\footnote{Here again the `accidental' value $n=4$ where the scalar goes away produces the 4D partially massless field which is conformally invariant and also has a scalar gauge invariance.}. The transverse vector is accompanied by an additional $\Delta_L^{(1)}-4\lambda(n-1)$ factor that makes the operator of order four in derivatives; whereas the scalar is accompanied by additional $\Delta_L^{(0)}$ and $\Delta_L^{(0)}-2\lambda n$ factors that make the operator of order six in derivatives. The functional determinants of these additional factors cancel out against the Jacobian of the splitting and one gets the identity below that connects functional determinants for non-minimal operators with those for minimal (Laplace-like) ones  
\begin{equation}
\displaystyle
{\det}_{_{\top}}\,P_{2}^{(2)}\,=\,{\det}_{_{\bot\top}}\,P_{2}^{(2)}\cdot{\det}_{_{\bot}}\,P_{2}^{(1)}\cdot{\det}\,P_{2}^{(0)}~.
\end{equation}
The decoupling carries on for generic $k$. The non-minimal operators on an Einstein manifold, inferred from the conformally flat case\cite{Grigoriev:2018mkp}, are now given by the product
\begin{equation}
\displaystyle P_{2k}^{(2)}=\prod_{j=1}^{k}\left\{\Delta_L^{(2)}+\frac{2n}{(\frac{n}{2}-j+1)(\frac{n}{2}+j)}\,\nabla\,\nabla\cdot\,-\,4(n-1)\lambda +2\left(\frac{n}{2}-j\right)\left(j+\frac{n}{2}-1\right)\lambda\right\}~.
\end{equation}
Each factor of the above product, say j-block, decouples into j-blocks of the corresponding GJMS factors
\begin{eqnarray}
\displaystyle 
&\varphi_{_{\bot\top}}^{\mu\nu}\left\{\Delta_L^{(2)}\,-\,4(n-1)\lambda +2\left(\frac{n}{2}-j\right)\left(j+\frac{n}{2}-1\right)\lambda\right\}\varphi_{_{\bot\top},\mu\nu}\,+\\\nonumber\\\nonumber
&+2\frac{(\frac{n}{2}-j)(\frac{n}{2}+j-1)}{(\frac{n}{2}-j+1)(\frac{n}{2}+j)}V^{\bot,\,\mu}\left\{\Delta_L^{(1)}-4\lambda(n-1)\right\}\left\{\Delta_L^{(1)}-2(n-2)\lambda +2\left(\frac{n}{2}-j\right)\left(j+\frac{n}{2}-1\right)\lambda\right\}V^{\bot}_{\mu}+\\ \nonumber
\\
\nonumber
&+\frac{n-1}{n}\left[1-2\frac{n-1}{(\frac{n}{2}-j+1)(\frac{n}{2}+j)}\right]\,\sigma\,\Delta_L^{(0)}\left\{\Delta_L^{(0)}-2n\lambda\right\}\left\{\Delta_L^{(0)}+2\left(\frac{n}{2}-j\right)\left(j+\frac{n}{2}-1\right)\lambda\right\}\sigma~.
\end{eqnarray}
The functional determinant of the additional factors $\left\{\Delta_L^{(1)}-4\lambda(n-1)\right\}\,\Delta_L^{(0)}\,\left\{\Delta_L^{(0)}-2n\lambda\right\}$ again cancels out against the Jacobian, so that for generic order $k$ we simply obtain 
\begin{equation}
\displaystyle
{\det}_{_{\top}}\,P_{2k}^{(2)}\,=\,{\det}_{_{\bot\top}}\,P_{2k}^{(2)}\cdot{\det}_{_{\bot}}\,P_{2k}^{(1)}\cdot{\det}\,P_{2k}^{(0)}~.
\end{equation}

\section{One loop: central charges}

The holographic computation of the Weyl anomaly coefficients (or central charges) can be easily adapted form the massless cases that were already worked out in \cite{Aros_2019JHEP, Acevedo:2017vkk}. The two key ingredients are: (i) WKB exactness of the heat kernel when evaluated on the Poincar\'e-Einstein metric, and (ii) separation of curvature invariants into pure-Ricci terms that contribute to the volume anomaly and, in consequence, to the boundary Q-curvature and bulk pointwise Weyl invariants that contribute to the boundary ones depending on the dimension  \cite{Bugini:2016nvn}.
The boundary computation of the central charges, on the other hand, can be achieved by computing the accumulated heat kernel coefficients of the Lichnerowicz Laplacians on the boundary Einstein manifold \cite{Bastianelli:2000hi,Liu:2017ruz}. In this way, the 4D central charges can be confirmed in~\cite{Beccaria:2014xda} with appropriate conformal weights, while the charges for the 6D vector agree with those reported in~\cite{Beccaria_2017} (eqn. C7, therein).
We were able to match boundary and bulk calculations of all central charges, below we highlight the crucial steps in the holographic derivation that turns out to be much simpler and direct, while the boundary results for the individual GJMS-like operators are collected in \ref{CentralCharges}.\\

We rewrite the 4D Weyl anomaly in the basis of the Q-curvature and the pointwise Weyl invariant $W^2$ that is better suited for the holographic recipe:
\begin{eqnarray}
\mathcal{A}_4 &=&-a\,E_4 \,+\,c\,W^2\\\nonumber
 \\
\nonumber
&=&-4 a\,{\mathcal Q}_4 \,+\,(c-a)\,W^2 \nonumber\\\nonumber
\end{eqnarray}
In the 6D case, the preferred basis is the one containing the Q-curvature, the two cubic Weyl contractions $I_1$, $I_2$ and the Fefferman-Graham invariant $\Phi_6$:  \begin{eqnarray}
{\mathcal A}_6&=&-{a}\,E_6\,+\,{c_1}\,I_1\,+\,{c_2}\,I_2\,+\,{c_3}\,I_3 \\
\nonumber\\\nonumber
\qquad\quad&=&-48\,a\,{\mathcal Q}_6+(c_1+16c_3+32a)I_1+(c_2-4c_3-56a)I_2+(3c_3+24a)\Phi_6
\end{eqnarray}
Below we report directly the Weyl anomaly coefficients that come out in this preferred basis.
\subsection{5D bulk vector}
As mentioned before, the holographic computation can be easily adapted, it merely requires a shift in the mass-squared that brings in an exponential factor $e^{-k^2\,t}$. From Eqn. 2.8 in \cite{Acevedo:2017vkk}, we obtain the proper-time representation for the bulk functional determinant
\begin{align}
&\int_{0}^{\infty}\frac{dt}{t}\mbox{tr}_{_{\bot}}\,e^{-\left\{\hat{\Delta}_L^{(1)}-1+k^2\right\}\,t}
\\\nonumber
\\\nonumber
\sim&\int_{0}^{\infty}\frac{dt}{t^{7/2}}e^{-k^2t} \left[4\,+\,\frac{32}{3}t\,-\,\frac{11}{180}t^2\,\hat{W}^2 + \ldots  \right]. \nonumber
\end{align}
Taking the proper-time integral in terms of the gamma function, we read off the holographic Weyl anomaly coefficients
\begin{eqnarray}
a&=&\frac{1}{9}k^3 - \frac{1}{60}k^5\\\nonumber
\\
c-a&=&-\frac{11}{180}k
\end{eqnarray}
\subsection{5D bulk 2-tensor}
In the same fashion, from eqn.3.9 in \cite{Acevedo:2017vkk}, one gets for the 2-tensor

\begin{align}
&\int_{0}^{\infty}\frac{dt}{t}\mbox{tr}_{_{\bot\top}}\,e^{-\left\{\hat{\Delta}_L^{(2)}+4+k^2\right\}\,t}
\\\nonumber
\\\nonumber
\sim&\int_{0}^{\infty}\frac{dt}{t^{7/2}}e^{-k^2t} \left[9\,+\,54t\,+\,\frac{21}{20}t^2\,\hat{W}^2 + \ldots  \right]  \nonumber
\end{align}
that leads to
\begin{eqnarray}
a&=&\frac{9}{16}k^3 - \frac{3}{80}k^5\\\nonumber
\\
c-a&=&\frac{21}{20}k
\end{eqnarray}

A swift consistency check is given by the 2-tensor with $k=2$ in conjuction with the vector with $k=3$ : 4D Weyl graviton (e.g. eqn.1.6 in \cite{Tseytlin:2013jya}  or eqn.3.12 \cite{Acevedo:2017vkk})
$a_{_{Weyl}}=\frac{87}{20}$ and $c_{_{Weyl}}-a_{_{Weyl}}=\frac{137}{60}$.

\subsection{7D bulk vector}
Here we require the heat kernel in the 7D Poincar\'e-Einstein metric (eqn.3.3 and 3.12, \cite{Aros_2019JHEP}),

\begin{align}
&\int_{0}^{\infty}\frac{dt}{t}\mbox{tr}_{_{\bot}}\,e^{-\left\{\hat{\Delta}_L^{(1)}-4+k^2\right\}\,t}
\\\nonumber
\\\nonumber
\sim&\int_{0}^{\infty}\frac{dt}{t^{9/2}}e^{-k^2t} \left\{6\,+\,24t\,+\,\frac{72}{5}t^2\,-\,252\,\hat{W}^2\,\frac{t^2}{7!}\,-\,\left[\frac{472}{3}\hat{W}'^3\,-\,\frac{40}{3}\hat{W}^3\,+\,30\hat{\Phi}_7\,\right]\,\frac{t^3}{7!}\, + \ldots  \right\}  \nonumber
\end{align}
As part of the holographic recipe, we need to express the $\hat{W}^2$ term on the basis of Weyl invariants
that descend directly  to the boundary Weyl invariants of the anomaly: $\hat{W}^2=\hat{W}'^3-\frac{1}{4}\hat{W}^3+\frac{1}{4}\hat{\Phi}_7$. Then one reads off

\begin{eqnarray}
7!\cdot a&=&-\frac{1}{8}k^7 + \frac{7}{4}k^5-\frac{21}{8}k^3\\\nonumber
\\
7!\cdot(c_1+16\cdot c_3+32\cdot a)&=& 168 k^3 - \frac{472}{3}k\\\nonumber
\\
7!\cdot(c_2 - 4 \cdot c_3 - 56 \cdot a) &=& -42 k^3 + \frac{40}{3}k\\\nonumber
\\
7!\cdot(3 \cdot c_3 + 24\cdot a)&=& 42 k^3 - 30 k
\end{eqnarray}
This is in full agreement with the values reported in Eqn. C7 of \cite{Beccaria:2017dmw}.

\subsection{7D bulk 2-tensor}
Finally, from eqn.3.3 and 3.17 in \cite{Aros_2019JHEP} one finds

\begin{align}
&\int_{0}^{\infty}\frac{dt}{t}\mbox{tr}_{_{\bot\top}}\,e^{-\left\{\hat{\Delta}_L^{(2)}-1+k^2\right\}\,t}
\\\nonumber
\\\nonumber
\sim&\int_{0}^{\infty}\frac{dt}{t^{9/2}}e^{-k^2t} \left\{20\,+\,136t\,+\,\frac{256}{3}t^2\,+\,4760\,\hat{W}^2\,\frac{t^2}{7!}\right. \\ & \left. -\,\left[\frac{6064}{9}\hat{W}'^3\,+\,\frac{12368}{9}\hat{W}^3\,-\,348\hat{\Phi}_7\,\right]\,\frac{t^3}{7!} + \ldots  \right\}  \nonumber
\end{align}
that results in
\begin{eqnarray}
7!\cdot a &=&-\frac{5}{12}k^7 + \frac{119}{12}k^5-\frac{140}{9}k^3\\ \nonumber
\\
7!\cdot(c_1+16\cdot c_3+32\cdot a)&=& -\frac{9520}{3} k^3 - \frac{6064}{9} k \\ \nonumber
\\
7!\cdot(c_2 - 4 \cdot c_3 - 56 \cdot a) &=& \frac{2380}{3} k^3 - \frac{12368}{9} k\\\nonumber
\\
7!\cdot(3 \cdot c_3 + 24\cdot a)&=& -\frac{2380}{3} k^3 + 348 k
\end{eqnarray}

\section{Conclusion and outlook}

We have succeeded in extending the tree and one-loop holographic dictionaries to the family of GJMS-like operators acting on vector and symmetric 2-tensor fields. The holographic formula for the functional determinants seems to be quite useful as a building block in constructing partition functions and unveils a simple structure previously hidden from view.

There are several aspects of the present computation that deserve further studies, such as the transition to massless and partially massless bulk fields and the difficulties posed by the gauge symmetry to the extension to higher spins. In fact, for spin three and higher the Weyl tensor becomes an obstruction to the decoupling of transverse and longitudinal components.

Even though the factorization of functional determinants does not affect the computation of the central charges, the potential existence of a multiplicative anomaly may well affect the Casimir energy and the entanglement entropy.

\ack
We would like to thank Y. Matsumoto and A. Tseytlin for some useful conversations. This work was partially funded through FONDECYT-Chile 1220335.
\appendix

\section{Central charges}\label{CentralCharges}
Here we collect the boundary results by computing the accumulated heat coefficient $b_4$ and $b_6$ in 4D and 6D, respectively. We also include, for completeness, the central charges for the original GJMS operators for they are needed in order to compare with the holographic counterpart for the vector and 2-tensor, as follows from the holographic formula.\\

For computations using heat kernel coefficients, it is again better to rewrite the 4D Weyl anomaly in the basis of the Q-curvature and the pointwise Weyl invariant $W^2$; whereas in the 6D case, the preferred basis is now the one containing the Q-curvature, the two cubic Weyl contractions $I_1$, $I_2$ and the third pointwise Weyl invariant $I_3$:
\begin{eqnarray}
{\mathcal A}_6&=&-{a}\,E_6\,+\,{c_1}\,I_1\,+\,{c_2}\,I_2\,+\,{c_3}\,I_3 \\
\nonumber\\\nonumber
\qquad\quad&=&-48\,a\,{\mathcal Q}_6+(c_1-96a)I_1+(c_2-24a)I_2+(c_3+8a)I_3
\end{eqnarray}
Below we report directly the Weyl anomaly coefficients that come out in this preferred basis.\\

\begin{description}

\item[4D scalar $P_{2k}^{(0)}$]
\begin{eqnarray}
a&=&\frac{1}{144}k^3 - \frac{1}{240}k^5\\\nonumber
\\
c-a&=& \frac{1}{180}k
\end{eqnarray}
\item[4D vector $P_{2k}^{(1)}$]
\begin{eqnarray}
a&=&\frac{5}{48}k^3 - \frac{1}{80}k^5\\\nonumber
\\
c-a&=&-\frac{1}{15}k
\end{eqnarray}

\item[4D symmetric 2-tensors $P_{2k}^{(2)}$]

\begin{eqnarray}
a&=&\frac{65}{144}k^3 - \frac{1}{48}k^5\\\nonumber
\\
c-a&=&-\frac{10}{9}k
\end{eqnarray}

\item[6D scalar $P_{2k}^{(0)}$]
\begin{eqnarray}
7!\cdot a&=&-\frac{1}{48}k^7 + \frac{7}{48}k^5-\frac{7}{36}k^3\\\nonumber
\\
7!\cdot(c_1-96\cdot a)&=& \frac{56}{9}k^3 - \frac{80}{9}k\\\nonumber
\\
7!\cdot(c_2-24\cdot a)&=&-\frac{14}{9}k^3 + \frac{44}{9}k\\\nonumber
\\
7!\cdot(c_3+8\cdot a)&=&-\frac{14}{9}k^3 + 3k
\end{eqnarray}

\item[6D vector $P_{2k}^{(1)}$]
\begin{eqnarray}
7!\cdot a&=&-\frac{5}{48}k^7 + \frac{77}{48}k^5-\frac{175}{72}k^3\\\nonumber
\\
7!\cdot(c_1-96\cdot a)&=&-\frac{560}{9}k^3 + \frac{104}{9}k\\\nonumber
\\
7!\cdot(c_2-24\cdot a)&=&\frac{140}{9}k^3 - \frac{284}{9}k\\\nonumber
\\
7!\cdot(c_3+8\cdot a)&=&\frac{140}{9}k^3 - 13k
\end{eqnarray}

\item[6D symmetric 2-tensors $P_{2k}^{(2)}$]

\begin{eqnarray}
7!\cdot a&=&-\frac{7}{24}k^7 + \frac{289}{36}k^5-\frac{931}{72}k^3\\\nonumber
\\
7!\cdot(c_1-96\cdot a)&=&\frac{10024}{9}k^3 - \frac{22792}{9}k\\\nonumber
\\
7!\cdot(c_2-24\cdot a)&=&-\frac{2506}{9}k^3 - \frac{79522}{9}k\\\nonumber
\\
7!\cdot(c_3+8\cdot a)&=&-\frac{2506}{9}k^3 +126k
\end{eqnarray}

\end{description}

\section*{References}


\providecommand{\href}[2]{#2}\begingroup\raggedright\endgroup

\end{document}